\documentclass[useAMS,usenatbib]{mn2e_v2}
\usepackage{amsmath,graphics,epsfig}

\title[Waveguide-to-planar circuit transition for millimetre-wave detectors]{Waveguide-to-planar circuit transition for millimetre-wave detectors}
\author[G. Yassin, P.K. Grimes, O.G. King, and C.E. North]{G. Yassin\thanks{}, P.K. Grimes, O.G. King, and C.E. North\\
Dept. of Physics, University of Oxford, Denys Wilkinson Building, Keble Road, Oxford OX1 3RH, UK\\}

\begin{document}
\date{\today}
\pagerange{\pageref{firstpage}--\pageref{lastpage}} \pubyear{2008}
\maketitle

\begin{abstract}
We present a novel design of a waveguide to microstrip or coplanar waveguide transition using a unilateral finline taper. The transition from the unilateral finline mode to the TEM microstrip mode is done directly, avoiding the antipodal finline tapers that have commonly been employed. This results in significant simplification of the design and fabrication, and shortening of the chip length, thereby reducing insertion loss. In this paper we shall present designs at 90~GHz that can be employed in superconducting tunnel junction mixers or Transition Edge Sensor bolometers, and scale-model measurements at 15~GHz.
\end{abstract}

\footnotetext{Email: ghassan@astro.ox.ac.uk\\This paper is a preprint of a paper accepted by Electronics Letters and is subject to Institution of Engineering and Technology Copyright. When the final version is published, the copy of record will be available at IET Digital Library}

\section{Introduction}
\begin{figure}
\caption[Photograph of scale model replicas of the antipodal (top) and direct coupling to microstrip (bottom) transitions]{Photograph of scale model replicas of the antipodal (top) and direct coupling to microstrip (bottom) transitions}
 \centering
 \includegraphics[width=8.6cm]{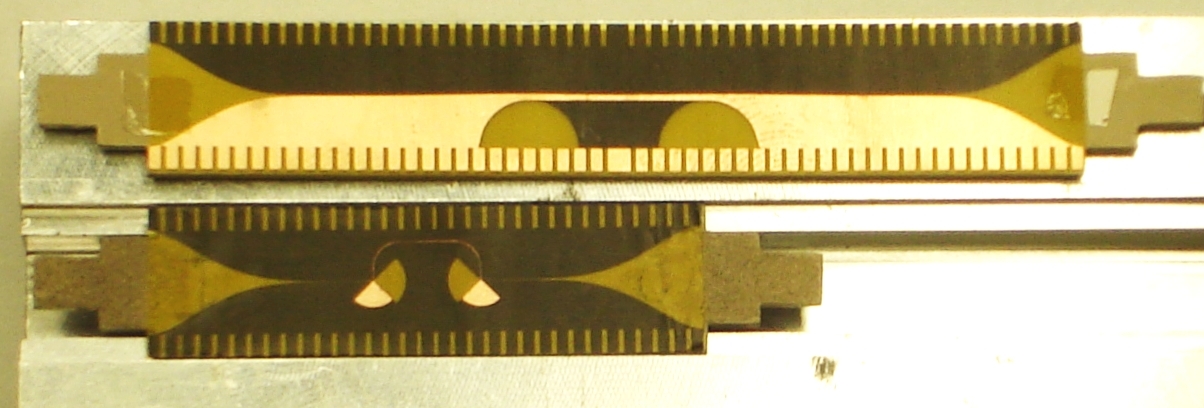}
 \label{fig:combined_pic}
\end{figure}

A high performance astronomical millimetre wave receiver consists of an array of horns which couples power from the sky to cryogenic detectors. The detectors are usually fabricated in superconducting planar circuits whose components are in most cases miniature microstrip lines, hence an efficient transition from waveguide to microstrip is needed. In previous publications \citep{Yassin:1997,Yassin:2000,Kittara:2004} we have reported the successful operation of SIS (Superconductor-Insulator-Superconductor) mixers at frequencies ranging from 220-700~GHz using antipodal finline tapers. The antipodal finline taper \citep{Yassin:2000} transforms the waveguide mode into the TEM microstrip mode using overlapping superconducting Nb films, separated by 400~nm of SiO oxide. The taper is deposited on a $\sim100$~$\mu$m quartz substrate which supports the structure in the E-plane of a rectangular waveguide. Before the fins overlap, the taper acts as a unilateral since the oxide is much thinner than the quartz substrate. When the fins start to overlap it behaves like an antipodal finline, and when the overlap becomes larger than the oxide thickness the transition to microstrip is performed using a semicircular taper (see Fig. \ref{fig:combined_pic}). The supporting substrate, usually quartz or silicon, has a relatively high dielectric constant; here the matching between the unloaded waveguide and waveguide loaded with substrate is important. Broad band matching can be achieved by a 2-step reduction in the substrate width, shown in the photographs in Fig. \ref{fig:combined_pic}. This transformer can be optimised to give return losses of 15-20~dB across a wide ($\>30$\%) band. As we shall see later, the performance of the taper is largely determined by the substrate mismatch.

Finline tapers have several advantages, including broad band operation and ease of detector block fabrication. The substrate dimensions are large relative to the microstrip width, which allows elegant integration of additional circuits of the receiver on a single chip. For example, a microstrip line with a 400~nm insulating layer of SiO2 and an impedance of approximately 20~$\Omega$ is 3~$\mu$m wide \citep{Yassin:1995}, much smaller than the width of the substrate. The length of the taper is at least one free space wavelength long, so important circuits such as bandpass filters, balanced and image separating mixer circuits can easily be integrated \citep{Kerr:1996}. Recently, this design was used in conjunction with Transition Edge Sensors (TES) and delivered a coupling efficiency well above 90\% \citep{Audley:2008}. It is however evident that the antipodal section with overlapping fins is difficult both to design and to fabricate, particularly when the lateral separation between the fins is very small, since at that point the field is significantly influenced by both the oxide that separates the fins and the supporting substrate. This makes the computation complicated and requires a large amount of memory. We have also learned that a lot of care is needed when fabricating the overlapping fin sections in order to avoid shorts between the very closely spaced fins as they begin to overlap.

\section{The new finline transition to microstrip}

\begin{figure}
\caption[(a) Schematic of the new waveguide to microstrip transition with direct coupling between the slotline and microstrip. (b) Schematic of the proposed waveguide to coplanar waveguide transition.]{(a) Schematic of the new waveguide to microstrip transition with direct coupling between the slotline and microstrip. (b) Schematic of the proposed waveguide to coplanar waveguide transition.}
 \centering
 \includegraphics[width=8.6cm]{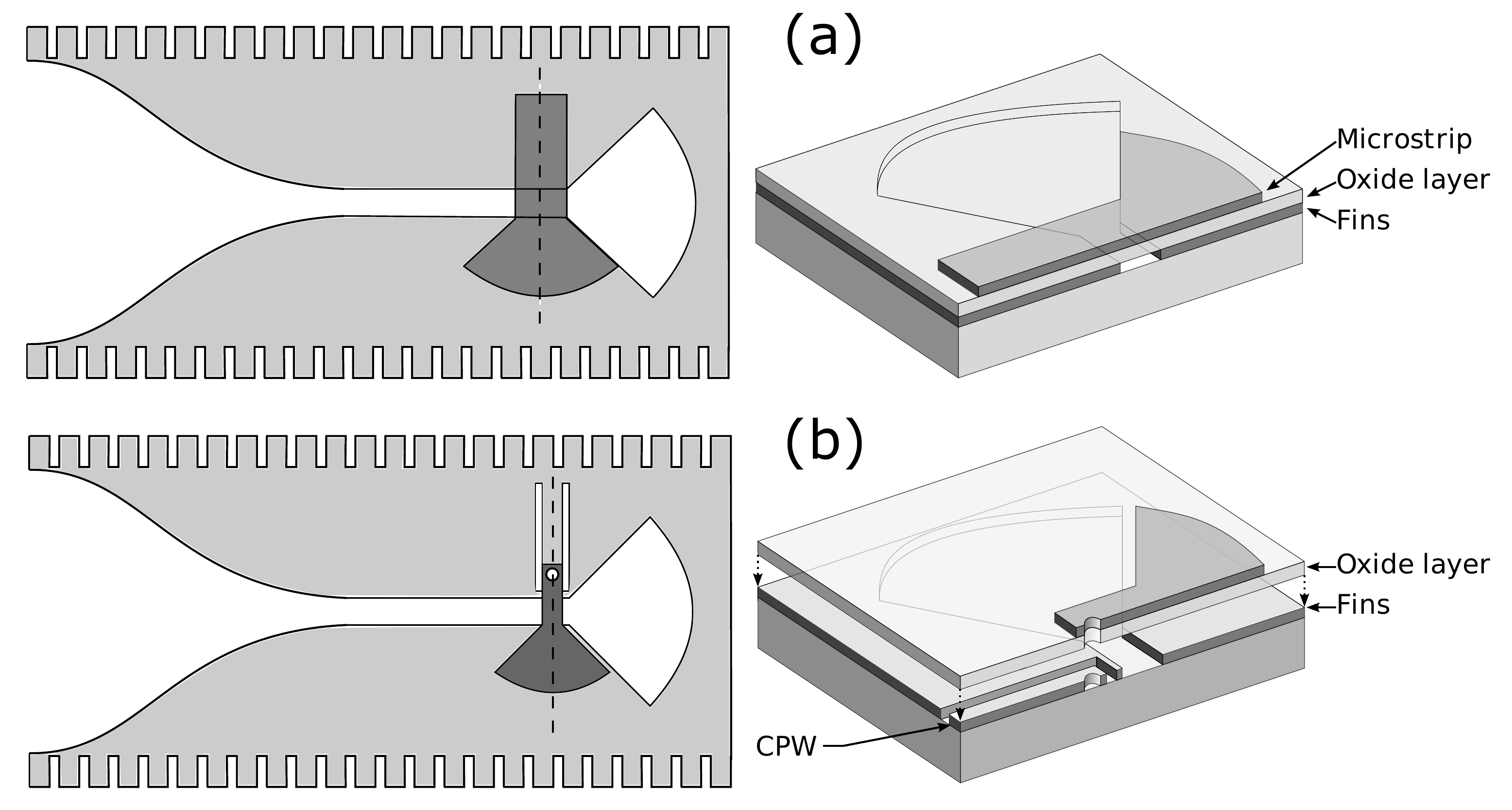}
 \label{fig:schematics}
\end{figure}

\begin{figure}
\caption[Results of HFSS simulations of slotline to microstrip transitions at 90 GHz, showing the return loss (RL) and insertion loss (IL).]{Results of HFSS simulations of slotline to microstrip transitions at 90 GHz, showing the return loss (RL) and insertion loss (IL).}
 \centering
 \includegraphics[width=8.6cm]{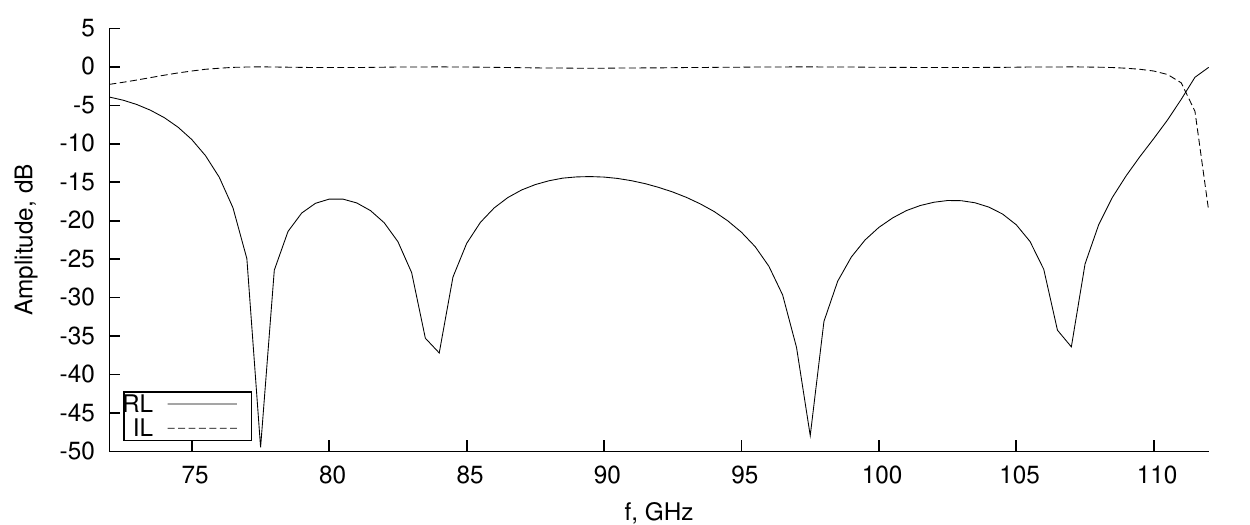}
 \label{fig:high_freq_sims}
\end{figure}

We shall now present a new design that retains the advantages mentioned above and yet makes the design and fabrication of the transformer much easier and significantly shorter. This is done by removing the antipodal section with overlapping fins and coupling the power directly from the slotline to the microstrip as shown in Fig.~\ref{fig:schematics}(a). As in the previous design, power couples from the waveguide to the unilateral finline taper, which is tapered to the desired impedance of the microstrip. For Nb films 400~nm thick deposited on 225~$\mu$m thick Si, an impedance of approximately 35~$\Omega$ is obtained when the gap becomes 3~$\mu$m. A microstrip bridge of the same impedance is deposited across the slotline on a 400~nm thick layer of Oxide (e.g. SiO2) and terminated by a shorted $\lambda$/4 radial stub. The finline itself is also terminated by a $\lambda$/4 radial stub which forms an RF short. We have simulated the performance of the new finline in HFSS at 90~GHz and shown that it exhibits excellent performance over 33\% bandwidth, which is sufficient for many applications such as Cosmic Microwave Background instruments. An example of these simulations is shown in Fig. \ref{fig:high_freq_sims}, which shows the return loss and insertion loss as a function of frequency for two back-to-back waveguide-to-microstrip transitions. These results show clearly that the performance of the transition is excellent over the required bandwidth.

\section{Transition to coplanar waveguide}

Transition from waveguide to coplanar waveguide (CPW) using the above mentioned method could be favoured over transition to microstrip in cases where higher transmission line impedances are required than are achievable using microstrips with convenient dimensions. For example, a 50~$\Omega$ microstrip line in a cryogenic detector using an oxide with a thickness of 400~nm and dielectric constant of 3.8 has a strip width of around 1~$\mu$m, requiring complicated lithography. This impedance, however, is ideal for coplanar waveguides since it can be achieved with a strip width of 5~$\mu$m and a gap of 2~$\mu$m, which is easily fabricated using photolithography along with the rest of the chip. The transition to CPW geometry is not significantly more complicated to fabricate than the transition to microstrip, requiring a via to connect the radial stub to the central conductor of the CPW, as shown schematically in Fig.~\ref{fig:schematics} (b).

\section{Scale model measurements}

We have designed and fabricated the two types of back-to-back transitions from finline to microstrip, optimised in the frequency band 11-16~GHz. The transitions were fabricated from Copper on RT/Duroid6010, which has a dielectric constant of 10.2 (similar to Silicon), with an Espanex insulating layer between the finline and microstrip layers. A photo of the transitions mounted in the aluminium split-block is shown in Fig.~\ref{fig:combined_pic}. The groove in the waveguide wall which supports the chip has a width of 2~mm in order to accommodate the serrations that cut-off the propagation of higher order modes in the WR10 waveguide. It can clearly be seen that the length of the new taper is much shorter than the conventional antipodal design.

\begin{figure}
\caption[Measurements and simulations of the substrate block alone in the waveguide, showing the return loss (RL) and insertion loss (IL).]{Measurements and simulations of the substrate block alone in the waveguide, showing the return loss (RL) and insertion loss (IL).}
 \centering
 \includegraphics[width=8.6cm]{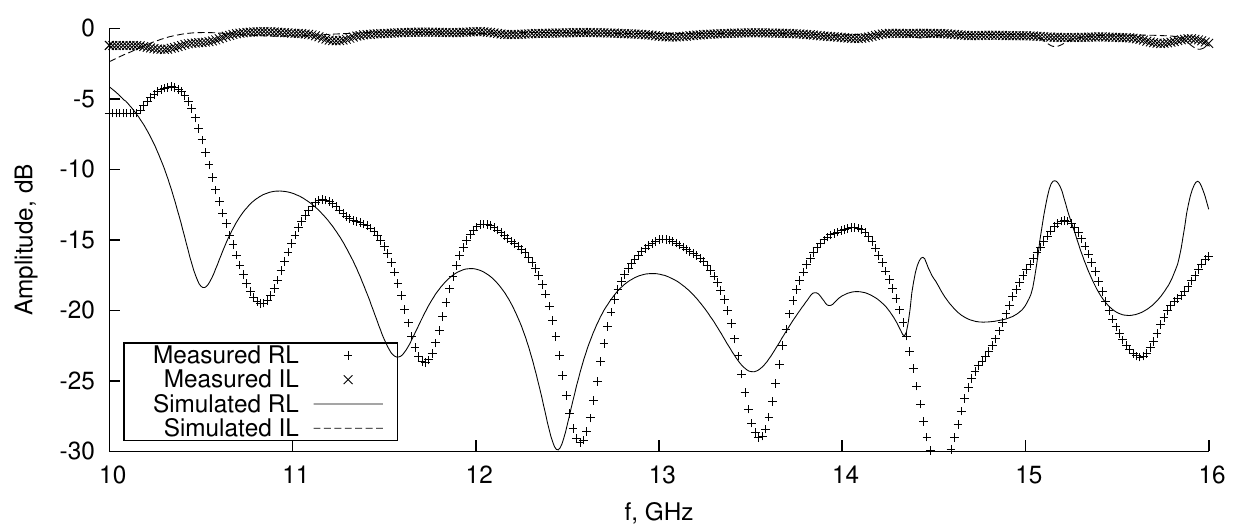}
 \label{fig:substrate_only_sims}
\end{figure}

\begin{figure}
\caption[Measurements and simulations of the new scale model transition from finline to microstrip, showing the return loss (RL) and insertion loss (IL).]{Measurements and simulations of the new scale model transition from finline to microstrip, showing the return loss (RL) and insertion loss (IL).}
 \centering
 \includegraphics[width=8.6cm]{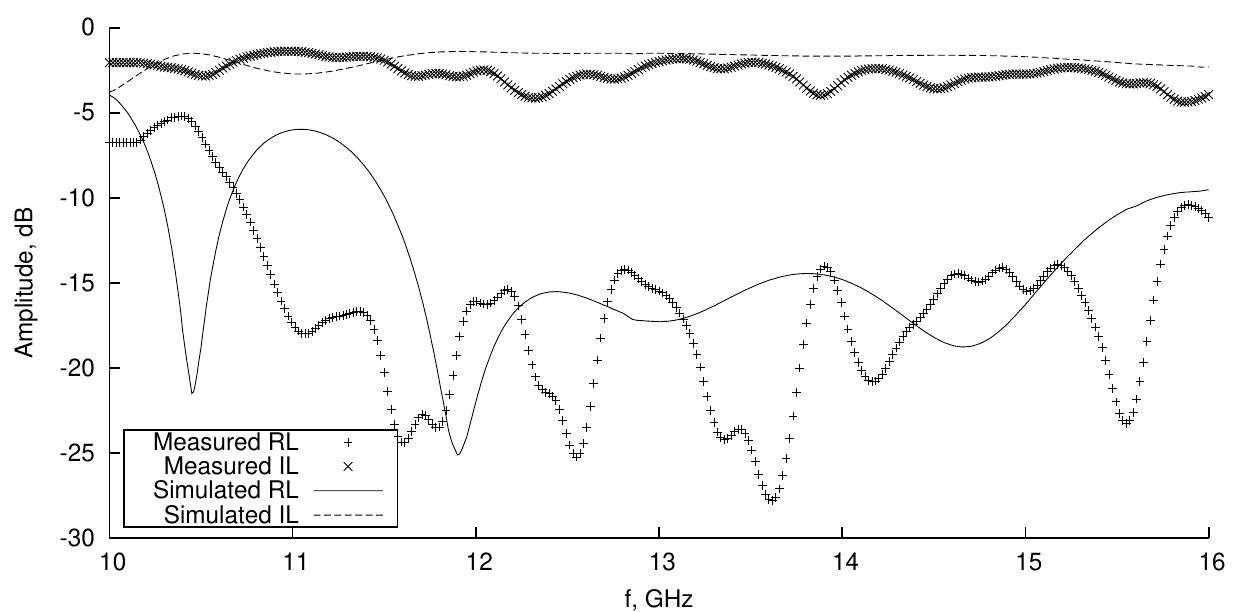}
 \label{fig:new_transition_sims}
\end{figure}

Measurements of the scattering parameters of scale models were made using a Vector Network Analyser and the scattering parameters of the substrate alone and the new transition are shown Figs.~\ref{fig:substrate_only_sims} and \ref{fig:new_transition_sims} respectively. There is in general very good agreement between the measurements and simulations, although at the low frequency end in Fig.~\ref{fig:new_transition_sims} the return loss simulations appear to have been slightly shifted (by approximately 0.5~GHz) with respect to the measured results. This could either be the result of inaccurate dielectric value of RT/Duroid6010LM (assumed 10.2) or fabrication tolerances. We also notice that the return loss in the useful bandwidth is largely determined by the 2-step substrate transition rather than the finline taper or the coupling to the microstrip, demonstrating the high quality of the transition design. Much better results should therefore be expected for thinner or lower dielectric constant substrate. The insertion loss is significantly influenced by conduction losses at room temperature which will not appear in real superconducting detector circuits. Finally we would like to draw attention to the fact that the scale model results are for two back-to-back tapers, so the return loss and insertion loss of a single taper should be reduced accordingly.

\section{Conclusions}

We have presented a new type of finline transition from waveguide to microstrip or to CPW. The transition is particularly suited to millimetre wave detector applications where the lateral dimensions become very close to the limits of what can be fabricated using standard photolithography. The transition to microstrip was greatly simplified by replacing the overlapping fins section with direct coupling from slotline to microstrip. In the case of transition to CPW, no additional layers of deposition are required. Scale model measurements agree well with simulated results.

\bibliographystyle{apalike_ru}
\pagestyle{plain} 
\bibliography{../../ogk.bib}

\begin{thebibliography}{}

\bibitem[Kerr and Pan, 1996]{Kerr:1996}
Kerr, A. and Pan, S.-K. (1996).
\newblock Design of planar image separating and balanced SIS mixers.
\newblock In {\em Proc. 7th Int. Symp. Space Terahertz Technol.}, pages
  207--219, Charlottesville, VA.

\bibitem[Kittara et~al., 2004]{Kittara:2004}
Kittara, P., Yassin, G., Withington, S., Jacobs, K., and Wulff, S. (2004), IEEE
  Trans. Microwave Theory Tech., 52, 2352--2360.

\bibitem[\mbox{Audley, M.D. et al.}, 2008]{Audley:2008}
\mbox{Audley, M.D. et al.} (2008).
\newblock Performance of microstrip-coupled TES bolometers with finline
  transitions.
\newblock In {\em Proceedings of the SPIE (in press)}, volume 7020.

\bibitem[Yassin et~al., 1997]{Yassin:1997}
Yassin, G., Padman, R., Withington, S., Jacobs, K., and Wulff, S. (1997),
  Electronics Letters, 33, 498--500.

\bibitem[Yassin and Withington, 1995]{Yassin:1995}
Yassin, G. and Withington, S. (1995), J. Phys. D: Appl. Phys., 28, 1983--1991.

\bibitem[Yassin et~al., 2000]{Yassin:2000}
Yassin, G., Withington, S., Jacobs, K., and Wulff, S. (2000), IEEE Trans.
  Microwave Theory Tech., 48, 662--669.

\end{thebibliography}

\end{document}